# Direct Growth of Graphene on Flexible Substrates toward Flexible Electronics: A Promising Perspective

Viet Phuong Pham

Additional information is available at the end of the chapter

**Abstract**

Graphene has recently been attracting considerable interest because of its exceptional conductivity, mechanical strength, thermal stability, etc. Graphene-based devices exhibit high potential for applications in flexible electronics, optoelectronics, and energy harvesting. In this paper, we review various growth strategies including metal-catalyzed transfer-free growth and direct-growth of graphene on flexible insulating substrates which are "major issues" for avoiding the complicated transfer process that cause graphene defects, residues, tears and performance degradation of its functional devices. Recent advances in practical applications based on "direct-grown graphene" are discussed. Finally, several important directions, challenges and perspectives in the commercialization of 'direct growth of graphene' are also discussed and addressed.

## 1. Introduction

Single-layer graphene (SLG) and few-layer graphene (FLG) films have been regarded as ideal materials for electronics and optoelectronics due to their excellent electrical properties and their ability to integrate with current top-down device fabrication technology [1–20]. Since the beginning of the twenty-first century, the interest in graphene materials has drastically increased, which is apparent in the number of annual publications on graphene (**Figure 1**). Until now, various strategies, including chemical vapor deposition (CVD) [21], liquid and mechanical exfoliation from graphite [16, 22, 23], epitaxial growth on crystal substrate [24–27], or solution-based processes on graphene oxides [28–34]. have been investigated for obtaining graphene layers. In particular, recent advances in CVD growth have successfully led to large-scale graphene production on metal substrates [1, 21, 35–41], driven by the high demand for utilizing graphene in possible applications of current complementary metal-oxide-semiconductor (CMOS) technology such as radio-frequency transistors, optical devices, and deposition processes [2].

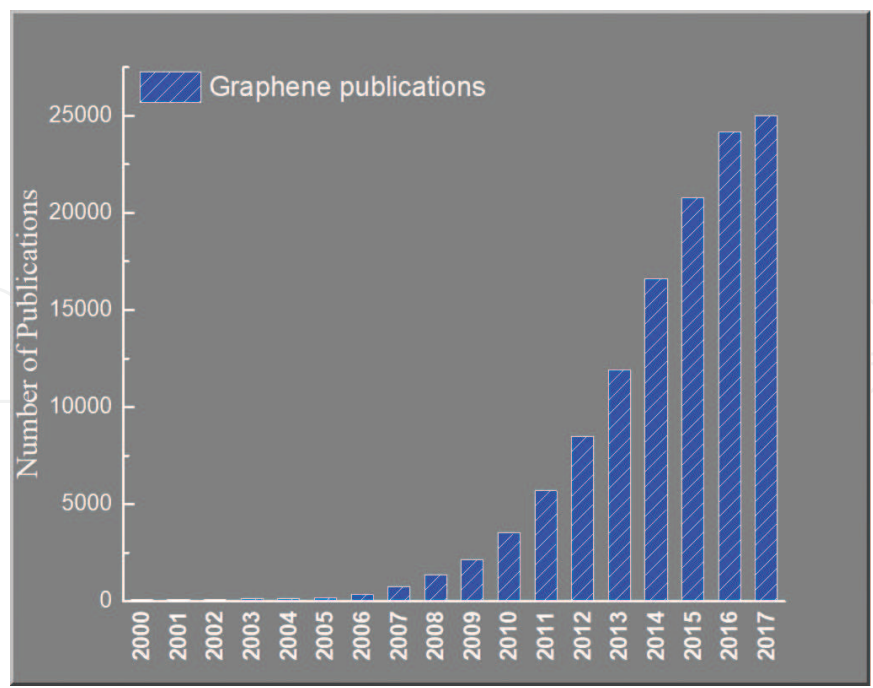

**Figure 1.** Publications on Graphene from 2000 to 2017. Source: ISI web of science (search: Topic = Graphene).

In these days, high-quality large-area graphene has been well synthesized on conducting metal substrates by using the catalytic CVD growth approach, which promoted a wide range of graphene-based device applications [1, 21, 35–42]. However, graphene grown on a metal substrate needs to be transferred onto dielectric substrates for electronic applications. Although various approaches, such as wet etching/transfer [38], mechanical exfoliation/transfer [16, 22, 23], bubbling transfer [43], electrochemical delamination [44–46], for transferring from the catalytic metal growth substrates to dielectric device substrates have been developed, none of these approaches is free from degradation of the transferred graphene. For example, 'wet etching and transfer', the most widely used transfer approach, is a serial process, which includes encapsulation the graphene surface with polymer supporting layer, subsequent chemical etching of the underlying metal substrate, transferring of the polymer/graphene film onto a dielectric substrate and removal of the polymer supporting layer [42]. Unfortunately, the transfer process is not just inconvenient but also causes various chemical and mechanical defects in the transferred graphene layer. CVD graphene grown on arbitrary substrates contains various defects, such as point defects, dislocation-like defects, cracks, wrinkles, and grain boundaries [47]. Because carbon atoms at such defect sites are chemically less stable than the carbon in defect-free graphene [47], the defect sites exposed to unavoidable surface/interface contaminants during the graphene transfer process are chemically damaged by bonding to oxygen, hydrogen, etc. Similarly, CVD graphene can be contaminated by metallic impurities from the growth substrate, which influences the electrochemical and electronic properties of graphene [48, 49]. CVD graphene never has 100% coverage and there are defects and holes which can be determined electrochemically [50]. In addition, transfer of ultrathin graphene layer on to target substrates leave unavoidable mechanical defects in the transferred graphene, such as cracks, tears and wrinkles. Thus, it leads to high deterioration in the performance of the resulting graphene-based devices, such as inducing a gradual reduction in the electrical conductivity of the devices or reducing the stability or increasing the leakage current of the devices.

Recently, significant efforts have been made to obtain graphene on semiconductor and dielectric substrates to avoid the problematic wet-transfer process. For example, a graphene was directly grown on a quartz substrate using a thin-layer Cu (100–450 nm thick) on the substrate as catalytic layer [51]. After growth on the Cu layer and the graphene layer could be transferred directly on the underlying dielectric surface through de-wetting and evaporating of Cu layer. Furthermore, the graphene on Cu was patterned for facile transfer to the underlying dielectric substrate after etching of the metal layer underneath the graphene [52]. However, the above transfer methods may only be suitable for small-size lateral graphene. Recently, Byun et al. have obtained FLG directly on $SiO_2$ using organic-polymer-coated insulators and thermal encapsulation of the Ni layer [53]. Lee et al. have observed the formation of FLG at the interfaces of Ni and $SiO_2$ using plasma-enhanced CVD (PECVD); however, the interface FLG was defective and thick [54]. The graphene formation on Ni is due to the dissociation and precipitation processes of the carbon species in Ni [54]. Consequently, the carbon precipitation is a non-equilibrium process and might be a major challenge for obtaining homogeneous graphene based on an Ni catalyst [36, 55]. The drawback of the metal catalyzed-CVD process is defects and tears unavoidably during the transfer process [56]. To avoid these drawbacks, two growth approaches have been suggested for the direct formation of graphene on flexible and rigid insulating substrates: (1) metal-catalyzed direct growth without transfer to external substrates [51], and (2) direct growth of graphene on a dielectric substrate without metal catalyst [57–59]. A recent report demonstrated graphene synthesized by metal-free CVD process on sapphire for forming large-scale highly crystalline SLG; however, it still has high wrinkles or ripples [60], similar to the situation in case of graphene pads/exfoliated h-BN [61]. Consequently, the metal-free CVD growth is not yet applicable to amorphous flexible and rigid substrates. A promising approach for increasing the graphene quality and minimizing the amount of metal catalyst is by evaporation and reaction with carbon-based gas precursors on the substrate surface. The formation of SLG and FLG on amorphous oxide substrates has been described in a previous report [62]. However, there are still many drawbacks that need to be addressed, particularly the amount of defects and non-uniform graphene due to imperfect nucleation and catalytic reactions. To the best of our understanding, there is no uniform high-quality SLG directly grown on dielectric substrates.

Considering that the mechanical transfer of graphene to the device substrates inevitably causes serious degradation in the performance of the resulting graphene devices, direct growth of graphene in a simple way on the flexible organic (e.g. polyimide (PI), PDMS, and Willow glass, mica) or rigid inorganic (e.g. glass, AlN, GaN, sapphire, quartz, Si, textured Si, $SiO_2$, SiC, fused silica, MgO, h-BN, $MnO_2$, $TiO_2$, and $HfO_2$) insulating substrates is highly desirable. The direct growth approach for device applications enables avoiding complex transfer process and transfer-induced defects. Moreover, it enables conformal growth of graphene on three-dimensional (3D) surfaces, which is necessary for various functional devices, such as sensors [63–65], black silicon solar cells [66], cambered micro-optics [67], 3D microelectromechanical system (MEMS) [68], or CMOS technology-based applications [2]. Here, we present an overview of various recently reported strategies for direct graphene growth on flexible substrates. In addition, a wide-range of applications as well as the perspectives and challenges are also addressed.

## 2. Generally growth mechanism of CVD-based graphene

CVD growth of graphene is a chemical process for the formation of SLG or FLG on an arbitrary substrate by exposing the substrate to the gas-phase precursors at controlled reaction conditions [69]. Owing to the versatile nature of CVD, intricately mixed homogeneous gas-phase and heterogeneous surface reactions are involved [70]. In general, as the partial pressure and/or temperature in the reaction substances are increased, homogeneous gas-phase reactions and the resulting homogeneous nucleation become significant [80]. To grow a high-quality graphene layer, this homogeneous nucleation needs be minimized [70]. A general mechanism for CVD-based graphene growth on catalytic metal substrates, for the growth of uniform and highly crystalline graphene layer on the surface, includes eight steps as follow: (1) mass transport of the reactant, (2) reaction of the film precursor, (3) diffusion of gas molecules, (4) adsorption of the precursor, (5) diffusion of the precursor into substrate, (6) surface reaction, (7) desorption of the product, and (8) removal of the by-product (**Figure 2**) [71].

Typically, CVD growth of 2D materials (e.g. graphene) involves catalytic activation of chemical reactions of precursors at the growth substrate surface/interface in a properly designed environment. Generally speaking, the roles of precursors, conditions (e.g. fast growth rates, large domain size, or very high crystalline quality), atmosphere, substrates and catalysts are the key factors affecting the final quality of the grown 2D materials. So far, significant efforts have been made to prepare highly crystalline 2D materials (e.g. graphene), but many challenges are still ahead. For example, due to the rough feature of catalytic metal surface, growth of uniform and high quality graphene is considerably difficult. The 2D material research community is also interested in new precursors (e.g. solid precursor only, gas precursor or solid precursor mixed with certain solvents) that could induce the formation of high-quality uniform graphene with minimal defect density. Another question is the effect of growth rate on the catalytic metal surface on the quality of graphene. Currently, it is difficult to give an exact answer, as investigations are progressing at an exponential rate.

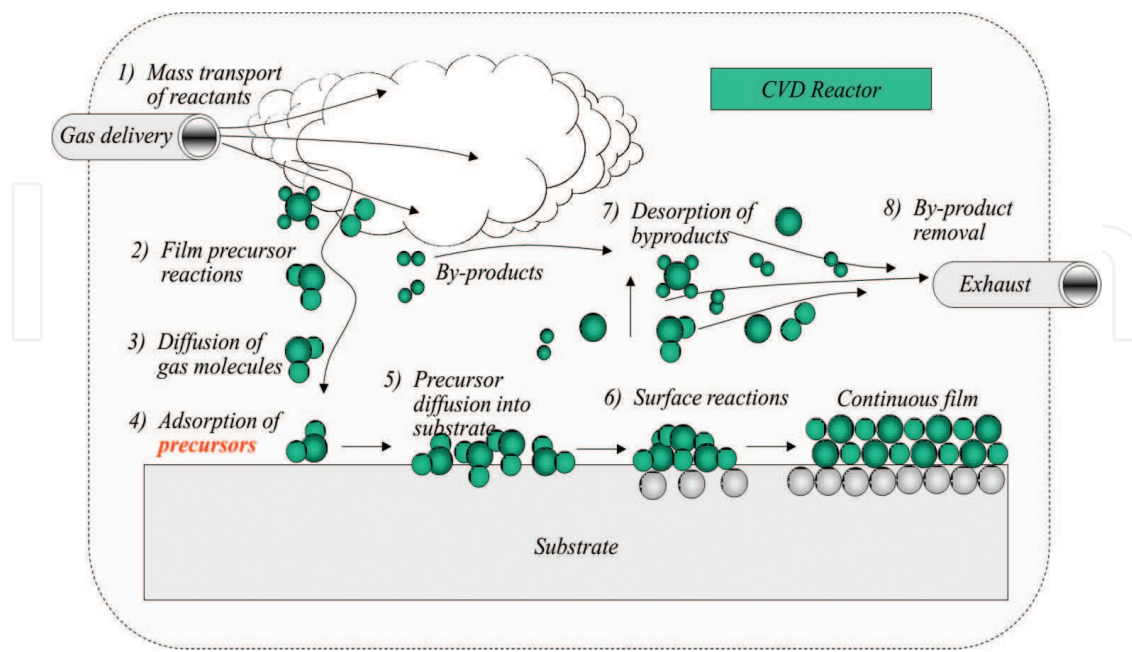

**Figure 2.** Diagram of generally growth mechanism of CVD-based graphene: Transport and reaction processes. Reproduced with permission from [71]. Copyright 2011, Freund Publishing.

However, non-catalytic direct-growth of graphene on semiconducting and dielectric rigid and flexible substrates follows different mechanisms according to our best insights. To date, the understanding of the concept of the general mechanism of the direct growth of graphene is still not yet adequate, neither experimentally nor theoretically, with many proposed possible growth mechanisms, e.g. vapor-solid-solid [72], or vapor-solid [73], or solid-liquid-solid [74]. There has been arguments on the direct-growth mechanism of graphene domains on dielectric rigid and flexible substrates or non-catalytic substrates in previous reports [59, 60, 75], but the mechanism for the entire process of the carbon precursor transformation to the crystalline graphene structures has not yet been fully understood. Thus, understanding the graphene growth mechanism and the effect of various growth conditions will be of significant interest to the 2D material research community to obtain large-scale, high-quality graphene.

## 3. Direct growth of graphene on flexible organic substrates at low temperature

To avoid the problems arising in the graphene transferring process, two growth approaches have been suggested for the direct formation of graphene on flexible insulating substrates without additional transfer processes: (i) catalytic growth with the help of an external metal catalyst [51], and (ii) non-catalytic direct growth of graphene on a dielectric substrate without a metal catalyst [57–59].

The direct growth of graphene is a process on flexible substrates (PI, PDMS and Willow glass, mica) and rigid substrates (glass, AlN, GaN, sapphire, quartz, Si, textured Si, $SiO_2$, SiC, fused silica, MgO, h-BN, $MnO_2$, $TiO_2$, and $HfO_2$) without transfer processes [6, 52–63, 74, 76–100], compared with conventional indirect growth processes on metal substrates (Cu, Ni, Ge, etc.) which need additional transfer processes onto arbitrary substrates [1–5, 8–51]. Using this method, we can avoid the complicated transfer process, which induces the defects, residues, and tears that degrade the performance of graphene-based devices. Various approaches for direct growth of graphene were classified into three major types: (i) catalyst-free and polymer-free, (ii) based on both catalyst and polymer, and (iii) based on metal catalyst (**Figure 3**).

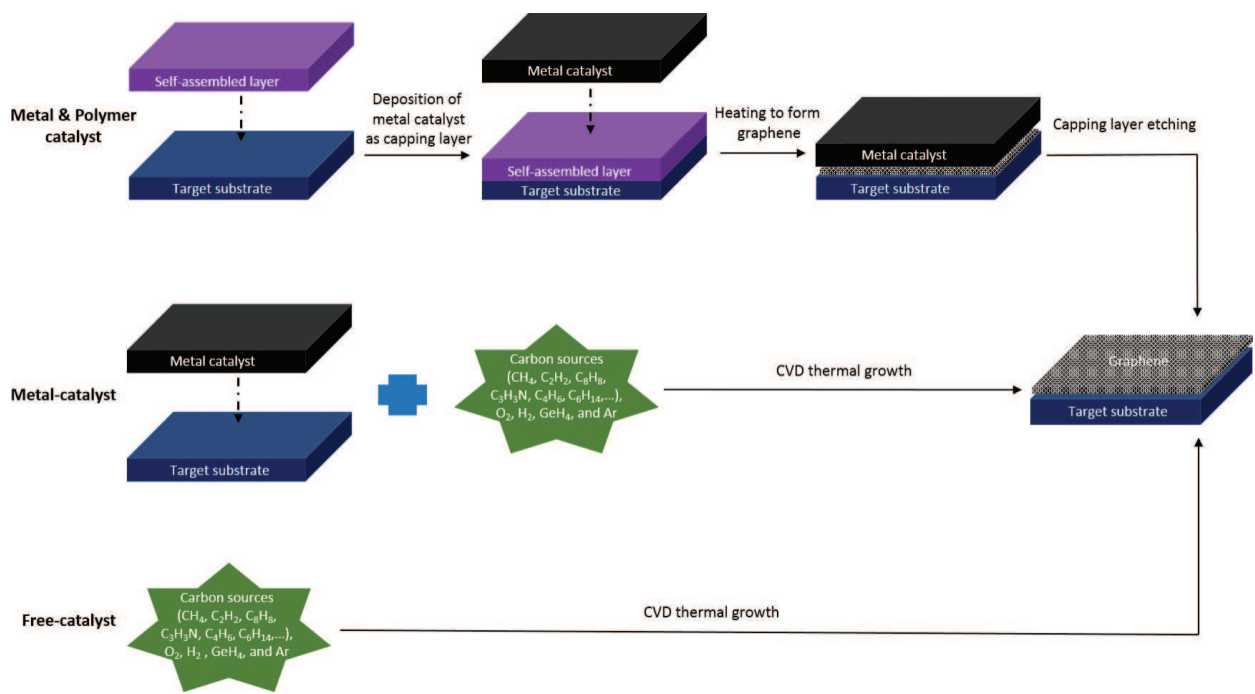

**Figure 3.** Generally methods for direct-growth of graphene onto target insulating substrate.

Compared with the conventional catalytic growth approach, the direct growth of graphene on dielectric substrate without any external catalysts has various advantages, such as low process cost, and the shorter experimental processes. However, the drawback of this method is that without using catalysts, the chemical reactions for the excitation of the kinetic energy of the graphene growth process is not sufficient to obtain high quality direct-grown graphene for commercialization, compared with the catalytic direct growth described in previous sections.

Direct growth of graphene on flexible organic substrates has a huge potential in applications related to flexible and stretchable electronics, such as e-skin and health monitoring on the human body [101–103]. However, the limited thermal stability of organic substrates, which can be easily melted, deformed or damaged at high temperatures (>300°C), leads to a serious limitation in direct growth of graphene on the flexible substrates, because the quality of graphene grown at low temperatures (<400°C) is much lower than that of the one at high temperatures (~1000°C). Owing to these constraints, graphene growth on flexible substrates is only recently being studied [77, 78]. To reduce the process temperature, most of the studied growth methods involve the catalytic conversion of organic precursors to graphitic layers on the flexible organic substrates with the help of catalytic metal layers.

In 2012, Kim et al. reported a low-temperature (300°C) growth of graphene-graphitic carbon (G-GC) films on Cu layer deposited on polyimide (PI) substrate using inductively coupled plasma-enhanced CVD (ICP-CVD), and a direct transfer of the G-GC films onto a underlying flexible PI substrate using wet etching of Cu layer (**Figure 4a–f**) [77]. The optical and electrical characteristics of G-GC are affected by the varying growth temperature, plasma power and growth time. More recently, in 2016, Seo et al. revealed a simple, inexpensive, scalable and patternable process to synthesize graphene-dielectric bi-layer (GDB) films on solution-processed polydimethylsiloxane (PDMS) under a Ni capping layer (**Figure 4g**) [78]. Seo et al. deposited the Ni film as the catalyst and encapsulation layer on a PDMS layer that was a few micrometer thick; this layer enabled direct growth of GDB between the substrate and Ni layer. PDMS (4 μm)/Ni (400 nm) films on the substrate were thermally annealed under vacuum, forming a PDMS/MLG/Ni/MLG structure. At the interface of the PDMS layer and the Ni film, the carbon atoms in the PDMS surface diffused into the Ni layer under high temperature, and carbon atoms were released to form MLG on both sides of the Ni layer during cooling. With this method the GDB structure was fabricated simultaneously and directly on the substrate, by thermal conversion of the PDMS without using additional graphene transfer and patterning process or formation of an expensive dielectric layer, which makes the device fabrication process much easier.

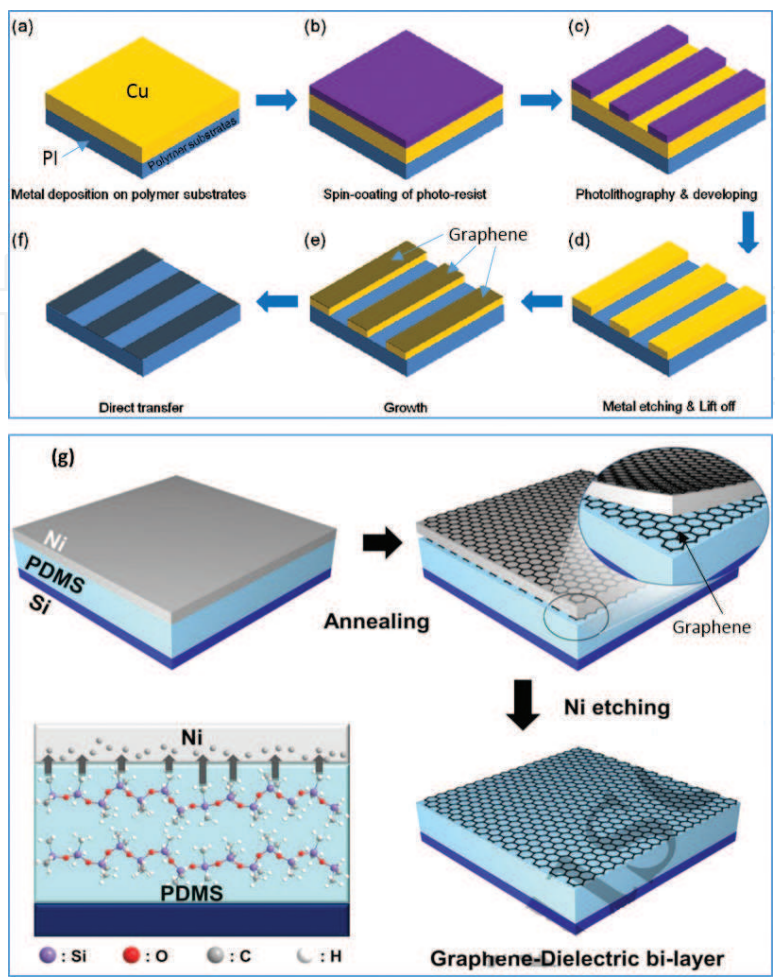

**Figure 4.** Diagram of direct-growth of graphene onto flexible substrates: (a–f) PI, and (g) PDMS. (a–f) Reproduced with permission from [77], copyright 2012, IOP publishing. (g) Diagram of direct-growth of bilayer graphene on PDMS substrate based on Ni catalyst (reproduced with permission from [78], copyright 2017, IOP Publishing).

In 2015, Sun et al. revealed a growth method of graphene-graphitic carbon (G-GC) at the growth conditions (low-temperature range 400–600°C, $CH_4$ gas, pressure 100 W, and growth time in 1 h) using PECVD system as depicted in **Figure 5** [108]. The advantage of direct PECVD process is that graphene films could be formed on flexible substrate, e.g.

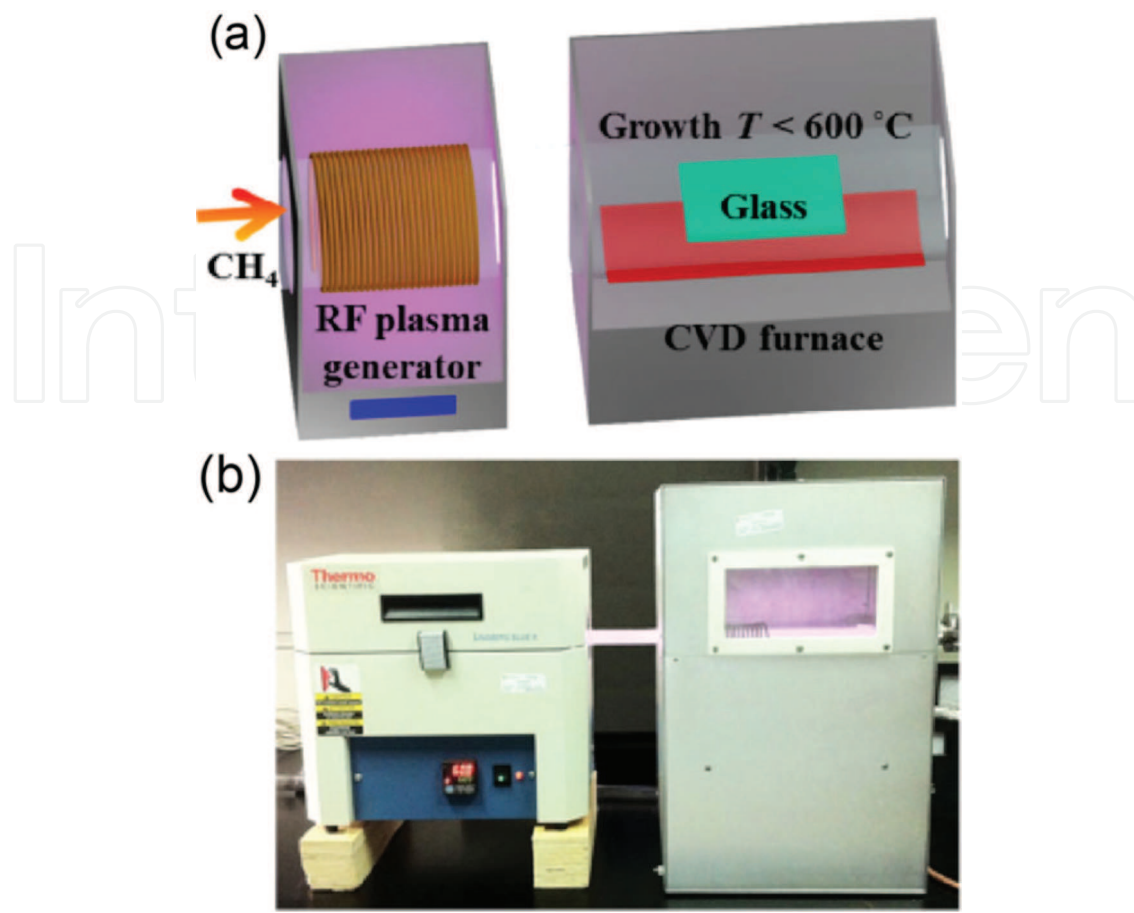

**Figure 5.** (a) Diagram of direct-growth of graphene onto flexible mica glass substrate: (b) PECVD system utilized in this graphene growth with a single-zone electrical chamber (left) and RF plasma source (right). (a, b) reproduced with permission from [100], copyright 2015, Springer and Tsinghua University Press.

mica substrate. The uniform and high-quality graphene films directly integrated with low-cost used flexible mica glass will unlock a promising perspective in fabrication of multi-functional electrodes in solar cell, smart window, and transparent electronic.

## 4. Application of direct-grown graphene on flexible electronics

A wide range of functional devices (transistors, solar cells, sensors, resistors, diffusion barriers, heat-resistant devices, photocatalytic plates and energy-saving smart windows) of graphene directly grown on various dielectric substrates using different growth methods, catalysts and device performances to date have been introduced, as briefly classified in **Table 1**.

### 4.1. Transistors (FETs)

The transfer-free growth of graphene will provide a new way to fabricate the FET-based electronic applications of graphene simply and inexpensively, while avoiding the transfer process. In general, FETs based on transfer-free direct graphene growth on various substrates (PDMS, sapphire, quartz, $SiO_2$ and h-BN), have been investigated thoroughly in previous studies [1, 6, 52, 59, 60, 62, 78, 82, 86, 88, 90, 92, 95, 97]. In particular, direct-grown graphene

| Substrate | Property | Method | Catalyst | Applications of direct-grown graphene | Results | Ref. |
|-----------|----------|--------|----------|----------------------------------------|---------|------|
| PI | Flexible | ICP-CVD | Cu | Strain sensor | Transmittance (%T) (77% at 550 nm) | [77] |
| | | | | | Sheet resistance (R$_s$) (80 KΩ/sq) | |
| PDMS | Flexible | Spin coat and thermal annealing, CVD | Ni | FET | Electron mobility μ$_e$ = 0.01 cm²/Vs | [78] |
| | | | | | On/off ratio (1.1 × 10⁴) | |
| Flexible Willow Glass | Flexible | CVD | Ni | Graphene pattern | High flexibility pattern | [98] |
| Mica | Flexible | PECVD | | Transparent circuit for LED | High transparent and flexibility | [100] |

Note: "NA" means "not applicable".

**Table 1.** A brief classification of direct-grown graphene on various flexible substrates and their applications to date.

on flexible PDMS-based FET device (**Figure 6**) will be a promising potential in future flexible electronics [78].

### 4.2. Strain sensor

Graphene can be used as 3D structured electrodes in multifunctional devices, such as pressure sensors [59], black silicon solar cells [66], cambered micro-optics [67], and MEMS sensors [68]. In general, graphene grown on catalytic metal substrate is almost impossible to be conformally transferred onto the 3D structural surface without mechanical damages [104]. Therefore, the direct growth of graphene on the 3D structured device surface can be a potential way to solve the limitations and problems above.

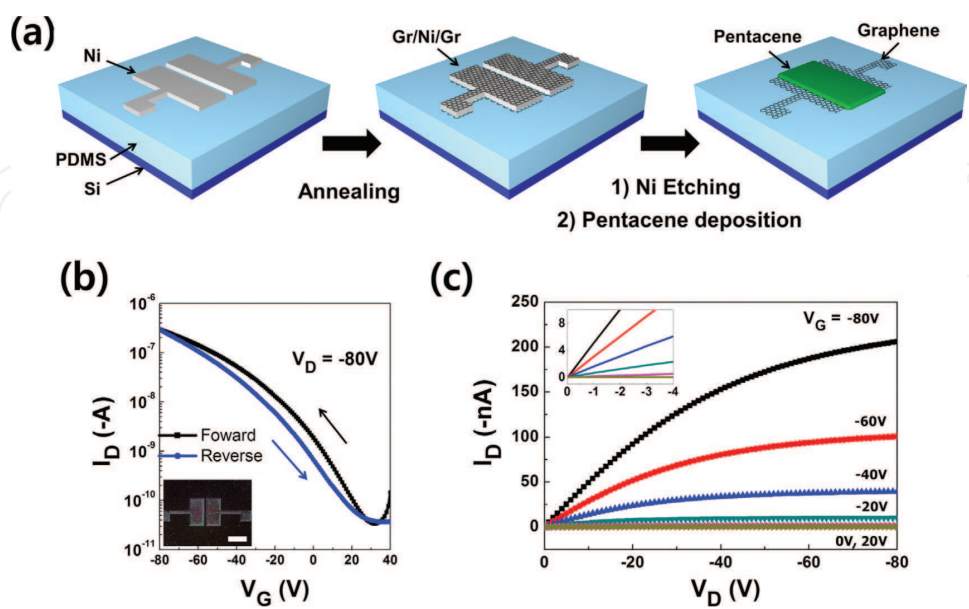

**Figure 6.** (a) Schematic of flexible PDMS-based FET device. (b) I-V curves of this FET device (black: Forward bias, blue: Reverse bias). Inset: Optical image of patterned source/drain graphene electrodes on a-PDMS. Scale bar: 1 mm. (c) Output characteristics of this FET device (channel length: 100 μm). Inset: Output characteristics at low voltage. (a-c) Reproduced with permission from [88], copyright 2017, IOP Publishing.

In 2012, by using CVD growth at low-temperature (300°C) for graphene films on a dielectric PI flexible substrate, Kim et al. successfully fabricated a graphene-based strain sensor on a PI substrate, and demonstrated the resistance modulation at different strains [77]. The resistance of graphene films showed a gradually increasing tensile strain of ~0.8% for 340 s (**Figure 7a**). Particularly, it linearly increased in the range of 31.64–31.69 MΩ with the applied strains of 0.1–0.8% (**Figure 7b**).

### 4.3. Strain pattern

To demonstrate the potential for applications of Ni-catalyzed direct-grown graphene on Willow flexible glass, Marchena et al. successfully produced ribbons and a square pattern of graphene (**Figure 8**) [98]. In particular, growing graphene directly on an ultrathin flexible Willow glass has huge potential in future flexible electronics.

### 4.4. Transparent circuit for green LED device

The directly grown graphene/glass sample, for the use in a range of transparent conductive applications (like transparent circuit) in industry, requires uniformity, low-cost, flexibility and good quality graphene on transparent flexible substrates (Corning Willow glass, or mica). It is currently in high demand to explore the novel functions of circuits on flexible glass, which may have application potentials in optoelectronics, gas/moisture/bio sensors etc. For instance, the resistances of the pattern graphene within circuit devices, would show a noticeable change without the harmful wet transfer graphene process applied directly on devices. In addition, regarding the potential for integration into flexible electronic devices, mechanical durability of the directly grown graphene is an important factor. To the best of our knowledge, such properties have not been studied so far.

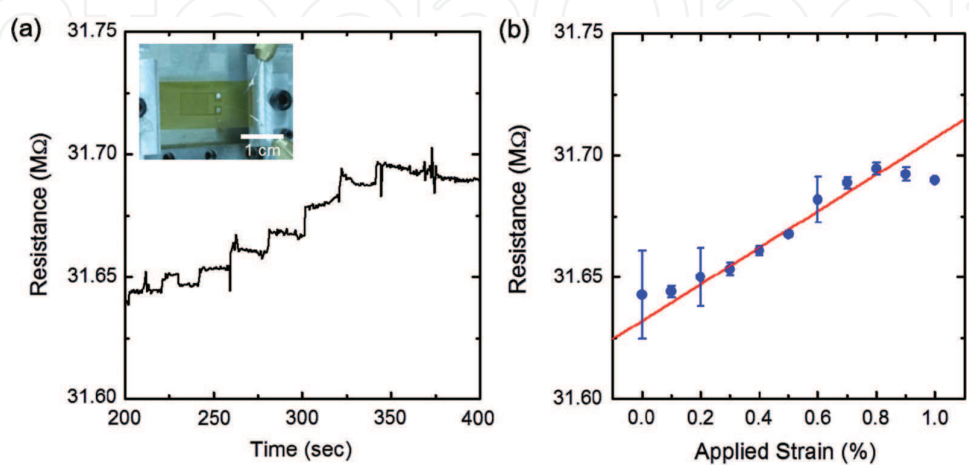

**Figure 7.** (a) Resistance changes of flexible strain sensor assisted direct-grown graphene on PI flexible substrate at various treatment time. The inset is strain-applied sensor. (b) Resistance changes of this device under applied strains. (a, b) reproduced with permission from [77], copyright 2012, IOP Publishing.

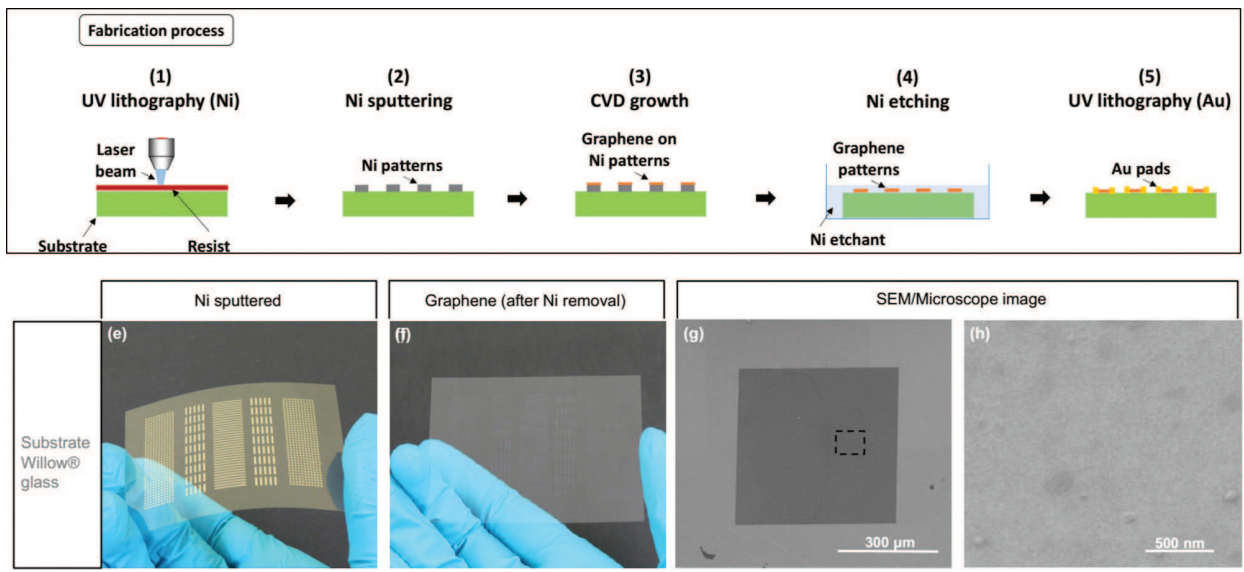

**Figure 8.** Sequences of direct-grown graphene pattern fabrication on willow flexible glass using Ni catalyst by UV lithography. Reproduced with permission from [98], copyright 2016, OSA Publishing.

Recently, Sun et al. fabricated and designed a transparent circuit based on PECVD direct-grown graphene on flexible mica glass sheet (2 × 8 cm) by utilizing photolithography, as shown in **Figure 9a** [100]. The results showed the patterned graphene electrode on a flexible mica glass can lighten up a green light-emitting diode (LED) indicator (**Figure 9b**). They also synthesized graphene directly on flexible mica glass and measured the change of the resistance through bending tests, with a bending variation of ~45%, and the full recovery after bending indicates good mechanical stability and flexibility of the graphene electrode (**Figure 9c**).

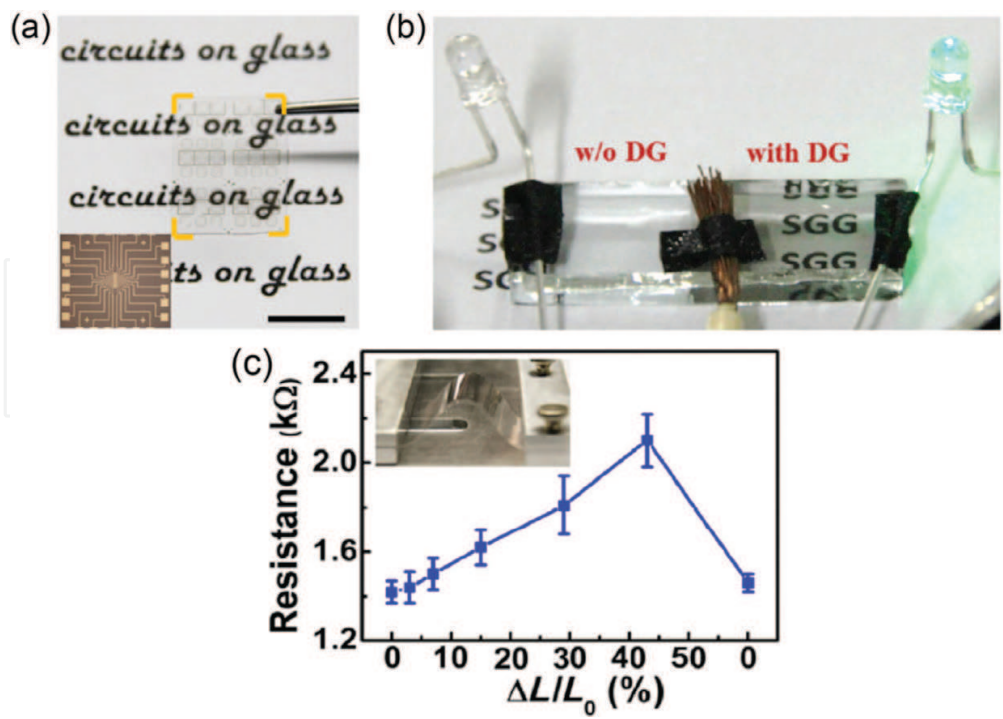

**Figure 9.** (a) Photograph of transparent circuit based on PECVD direct-grown graphene/flexible mica glass. The inset shows OM image of this device. (b) Photograph of patterned PECVD graphene on white glass showing the transparent conductivity to lighten up a green LED device. (c) Resistance with various bending values of direct-grown graphene film on mica glass. The inset shows the bending test. Reproduced with permission from [100], copyright 2015, Springer and Tsinghua University Press.

## 5. Conclusions, perspectives, and challenges

Strategies for direct graphene growth on arbitrary dielectric flexible substrates using the CVD method without metal catalyst at low temperature have been briefly reviewed. In addition, a wide range of device applications of the direct-grown graphene has been also discussed. The prospects of direct-grown transfer-free graphene are bright and currently receiving considerable attention from the 2D material research community. By discovering new methods for obtaining transfer-free graphene, the direct fabrication of a wide-range of various hetero-structured devices can be achieved. However, understanding the growth process and conditions that affect the quality of graphene is still very poor. So far, graphene grown directly on an insulating substrate is generally of low quality (**Table 1**). Because the direct growth relies on the thermal decomposition of carbon resources, the growth rate is usually low and size of the graphene domain is small, resulting in growth of defective graphene layer. Until now, many challenges remain in this direction and large-area high-quality graphene production is still very difficult. In order to obtain more advanced results, an in-depth understanding of the mechanism of graphene growth on insulating substrates is essential.

Direct graphene growth at low temperatures is an important research issue, because high growth temperatures are not allowed on many device substrates, such as flexible polymer and Si substrates. Direct growth of graphene at low- or near-room temperature [57, 77, 86], has been carried out. However, the results to date have not yet met the expectations. Given the practical application of graphene, several directions to pursue in direct graphene growth include low-temperature growth, high-speed growth of highly crystalline graphene, and direct growth on other two-dimensional materials (e.g. h-BN) located on flexible substrate (e.g. polyethylene terephthalate (PET), PDMS, PI, or mica). This issue also a very interesting issue toward future flexible electronic applications. The direct growth of large-scale graphene on h-BN is also an attractive topic [61, 83, 88–90, 105–107]. Graphene on h-BN can have excellent electrical properties, because h-BN is an ideal dielectric substrate for graphene devices, owing to its ultra-smooth, ultra-flat surface, insulation properties, chemical inertness and small lattice misfit compared to $SiO_2$ [111]. In theory, graphene synthesized could synthesize at low temperature on as-grown h-BN/flexible substrate exhibits better properties than that grown on transferred h-BN because of the transfer-induced contamination and defects on h-BN substrates. The transfer-free grown graphene on ultra-flat h-BN/flexible substrate, which could preserve the pristine properties of graphene, enables further promising flexible device applications based on vertically stacked 2D materials located on above flexible substrates.

Ultrafast direct growth of single-crystal graphene on flexible substrates is another fascinating and challenging topic, as ultrafast conventional indirect growth on copper has been investigated thoroughly and has progressed in recent years [108]. Further studies are required for obtaining faster direct growth and larger graphene single crystals on insulating flexible substrates. Several possible strategies are proposed: (i) to explore a more efficient way to reduce the reaction barrier for graphene direct growth; (ii) to obtain epitaxial, direct-grown well-aligned graphene domains and seamlessly stitch them together into a complete single-crystal film. One potential way to realize strategy (i) is probably to introduce a gas phase catalyst enhancing the catalytic conversion of carbon precursors to graphene layer. In fact, strategy (ii) has been implemented for conventional indirect-grown graphene on catalytic single-crystal substrates such as Ge (110) [5], or Cu (111) [109, 110]. However, the single-crystal graphene layer has limited size due to the limited single-crystal substrate size, and often the quality is less satisfactory owing to the imperfect alignment of individual graphene islands. Thus, to achieve a large-area, high-quality direct-grown graphene film, preparation of the large-area flexible substrate for releasing larger graphene nucleation seeds and an improvement in graphene alignment are critical issues in future works.

In addition, direct-growth of graphene on flexible substrate assisted by metal powder precursors (solid, diluted solution) contained inside a sub-chamber (as high temperature region) for a direct evaporation process into an innovative and re-designed-CVD main-chamber (containing flexible substrate as low temperature region) in order to allow graphene formation on dielectric substrates is an attractive topic and currently under investigation.

## Acknowledgements

This work is supported by the research and development grant, South Korea.

## Conflict of interest

There are no conflicts of interest to declare.

## Author details

Viet Phuong Pham

Address all correspondence to: pvphuong85@ibs.re.kr

SKKU Advanced Institute of Nano Technology, Sungkyunkwan University, Suwon, South Korea

## References

[1] Kim KS, Zhao Y, Jang H, Lee SY, Kim JM, Kim KS, Ahn JH, Kim P, Choi JY, Hong BH. Large-scale pattern growth of graphene films for stretchable. Nature. 2009;**457**:706-710. DOI: 10.1038/nature07719

[2]   Kim K, Choi JY, Kim T, Cho SH, Chung HJ. A role for graphene in silicon-based semicon-
      ductor devices. Nature. 2011;**479**:338-344. DOI: 10.1038/nature10680

[3]   Duong DL, Han GH, Lee SM, Gunes F, Kim ES, Kim ST, Kim H, Ta QH, So KP, Yoon
      SJ, Chae SJ, Jo YW, Park MH, Chae SH, Lim SC, Choi JY, Lee YH. Probing graphene
      grain boundaries with optical miscroscopy. Nature. 2012;**490**:235-239. DOI: 10.1038/
      nature11562

[4]   Choi JY. Graphene transfer: A stamp for all substrates. Nature Nanotechnology. 2013;**8**:311-
      312. DOI: 10.1038/nnano.2013.74

[5]   Lee JH, Lee EK, Joo WJ, Jang Y, Kim BS, Lim JY, Choi SH, Ahn SJ, Ahn JR, Park MH, Yang
      CW, Choi BL, Hwang SW, Whang D. Wafer-scale growth of single-crystal monolayer
      graphene on reusable hydrogen-terminated germanium. Science. 2014;**344**:286-289. DOI:
      10.1126/science.1252268

[6]   Shin HJ, Choi WM, Yoon SM, Han GH, Woo YS, Kim ES, Chae SJ, Li XS, Benayad A, Loc
      DD, Gunes F, Lee YH, Choi JY. Transfer-free growth of few-layer graphene by self-assem-
      bled monolayers. Advanced Materials. 2011;**23**:4392-4397. DOI: 10.1002/adma.201102526

[7]   Chae SJ, Güneş F, Kim KK, Kim ES, Han GH, Kim SM, Shin H, Yoon SM, Choi JY, Park
      MH, Yang CW, Pribat D, Lee YH. Synthesis of large-area graphene layers on poly-
      nickel substrate by chemical vapor deposition: Wrinkle formation. Advanced Materials.
      2009;**21**:2328-2333. DOI: 10.1002/adma.200803016

[8]   Güneş F, Shin HJ, Biswas C, Han GH, Kim ES, Chae SJ, Choi JY, Lee YH. Layer-by-layer
      doping of few-layer graphene film. ACS Nano. 2010;**4**:4595-4600. DOI: 10.1021/nn1008808

[9]   Pham VP, Jang HS, Whang D, Choi JY. Direct growth of graphene on rigid and flexible sub-
      strates: Progress, applications, and challenges. Chemical Society Reviews. 2017;**46**:6276-
      6300. DOI: 10.1039/c7cs00224f

[10]  Pham VP, Nguyen MT, Park JW, Kwak SS, Nguyen DHT, Mun MK, Phan HD, Kim
      DS, Kim KH, Lee J, Lee NE, Yeom GY. Chlorine-trapped CVD bilayer graphene for
      resistive pressure sensor with high detection limit and high sensitivity. 2D Materials.
      2017;**4**:025049. DOI: 10.1088/2053-1583/aa6390

[11]  Pham VP, Mishra A, Yeom GY. The enhancement of hall mobility and conductiv-
      ity of CVD graphene through radical doping and vacuum annealing. RSC Advances.
      2017;**7**:16104-16108. DOI: 10.1039/c7ra01330b

[12]  Pham VP, Kim DS, Kim KS, Park JW, Yang KC, Lee SH, Kim KN, Yeom GY. Low energy
      BCl3 plasma doping of few-layer graphene. Science of Advanced Materials. 2016;**8**:884-890.
      DOI: 10.1166/sam.2016.2549

[13]  Kim KN, Pham VP, Yeom GY. Chlorine radical doping of a few layer graphene with low
      damage. ECS Journal of Solid State Science and Technology. 2015;**4**:N5095-N5097. DOI:
      10.1149/2.0141506jss

[14]  Pham VP, Kim KN, Jeon MH, Kim KS, Yeom GY. Cyclic chlorine trap-doping for transparent,
      conductive, thermally stable and damage-free graphene. Nanoscale. 2014;**6**:15301-15308.
      DOI: 10.1039/c4nr04387a

[15] Pham VP, Kim KH, Jeon MH, Lee SH, Kim KN, Yeom GY. Low damage pre-doping on CVD graphene/cu using a chlorine inductively coupled plasma. Carbon. 2015;**95**:664-671. DOI: 10.1016/j.carbon.2015.08.070

[16] Geim AK, Novoselov KS. The rise of graphene. Nature Materials. 2007;**6**:183-191. DOI: 10.1038/nmat1849

[17] Neto AHC, Guinea F, Peres NMR, Novoselov KS, Geim AK. The electronic properties of graphene. Reviews of Modern Physics. 2009;**81**:109-162. DOI: 10.1103/RevModPhys.81.109

[18] Schwierz F. Graphene transistor. Nature Nanotechnology. 2010;**5**:487-496. DOI: 10.1038/nnano.2010.89

[19] Wilson NR, Macpherson JV. Carbon nanotube tips for atomic force microscopy. Nature Nanotechnology. 2009;**4**:483-491. DOI: 10.1038/nnano.2009.154

[20] Lin YM, Dimitrakopoulos C, Jenkins KA, Farmer DB, Chiu HY, Grill A, Avouris P. 100GHz transistor from wafer-scale epitaxial graphene. Science. 2010;**327**:662. DOI: 10.1126/science.1184289

[21] Yan Z, Peng Z, Tour JM. Chemical vapor deposition of graphene single crystals. Accounts of Chemical Research. 2014;**47**:1327-1337. DOI: 10.1021/ar4003043

[22] Su CY, Lu AY, Xu Y, Chen FR, Khlobystov AN, Li LJ. High-quality thin graphene films from fast electrochemical exfoliation. ACS Nano. 2011;**5**:2332-2339. DOI: 10.1021/nn200025p

[23] Hernandez Y, Nicolosi V, Lotya M, Blighe FM, Sun Z, De S, McGovern I, Holland B, Byrne M, Gun'Ko YK. High-yield production of graphene by liquid-phase exfoliation of graphite. Nature Nanotechnology. 2008;**3**:563-568. DOI: 10.1038/nnano.2008.215

[24] Sutter PW, Flege JI, Sutter EA. Epitaxial graphene on ruthenium. Nature Materials. 2008;**7**:406-411. DOI: 10.1038/nmat2166

[25] Berger C, Song Z, Li X, Wu X, Brown N, Naud C, Mayou D, Li T, Hass J, Marchenkov AN. Electronic confinement and coherence in patterned epitaxial graphene. Science. 2006;**312**:1191-1196. DOI: 10.1126/science.1125925

[26] Emtsev KV, Speck F, Seyller T, Ley L, Riley JD. Interaction, growth, and ordering of epitaxial graphene on SiC{0001} surfaces: A comparative photoelectron spectroscopy study. Physical Review B. 2008;**77**:155303. DOI: 10.1103/PhysRevB.77.155303

[27] Hass J, Heer WA, Conrad EH. The growth and morphology of epitaxial multilayer graphene. Journal of Physics. Condensed Matter. 2008;**20**:323202. DOI: 10.1088/0953-8984/20/32/323202

[28] Su CY, Xu Y, Zhang W, Zhao J, Liu A, Tang X, Tsai CH, Huang Y, Li LJ. Highly efficient restoration of graphitic structure in graphene oxide using alcohol vapors. ACS Nano. 2010;**4**:5285-5292. DOI: 10.1021/nn10169m

[29] Williams G, Seger B, Kamat PV. $TiO_2$-graphene nanocomposites UV-assisted photocatalytic reduction of graphene oxide. ACS Nano. 2008;**2**:1487-1491. DOI: 10.1021/nn800251f

[30] Green AA, Hersam MC. Solution phase production of graphene with controlled thickness via density differentiation. Nano Letters. 2009;**9**:4031-4036. DOI: 10.1021/nl902200b

[31]  Cote LJ, Kim F, Huang J. Langmuir-Blodgett assembly of graphite oxide single layers. Journal of the American Chemical Society. 2008;**131**:1043-1049. DOI: 10.1021/ja806262m

[32]  Li D, Müller MB, Gilje S, Kaner RB, Wallace GG. Processable aquaous dispersions of graphene nanosheets. Nature Nanotechnology. 2008;**3**:101-105. DOI: 10.1038/nnano.2007.451

[33]  Gao W, Alemany LB, Ci L, Ajayan PM. New insights into the structure and reduction of graphite oxide. Nature Chemistry. 2009;**1**:403-408. DOI: 10.1038/nchem.281

[34]  Joshi RK, Alwarappan S, Yoshimura M, Sahajwalla V. Graphene oxide: The new membrane material. Appl. Mater. Today. 2015;**1**:1-12. DOI: 10.1016/j.apmt.2015.06.002

[35]  Coraux J, Diaye ATN, Busse C, Michely T. Structural coherency of graphene on Ir(111). Nano Letters. 2008;**8**:565-570. DOI: 10.1021/nl0728874

[36]  Lee Y, Bae S, Jang H, Jang S, Zhu SE, Sim SH, Song YI, Hong BH, Ahn JH. Wafer-scale synthesis and transfer of graphene films. Nano Letters. 2010;**10**:490-493. DOI: 10.1021/nl903272n

[37]  Reina A, Jia X, Ho J, Nezich D, Son H, Bulovic V, Dresselhaus MS, Kong J. Large area, few-layer graphene films on arbitrary substrates by chemical vapor deposition. Nano Letters. 2008;**9**:30-35. DOI: 10.1021/nl801827v

[38]  Li X, Cai W, An J, Kim S, Nah J, Yang D, Piner R, Velamakanni A, Jung I, Tutuc E. Large-area synthesis of high-quality and uniform graphene films on copper foils. Science. 2009;**324**:1312-1314. DOI: 10.1126/science.1171245

[39]  Lee S, Lee K, Zhong Z. Wafer scale homogeneous bilayer graphene films by chemical vapor deposition. Nano Letters. 2010;**10**:4702-4707. DOI: 10.1021/nl1029978

[40]  Yan K, Peng H, Zhou Y, Li H, Liu Z. Formation of bilayer Bernal graphene: Layer-by-layer epitaxy via chemical vapor depostion. Nano Letters. 2011;**11**:1106-1110. DOI: 10.1021/nl104000b

[41]  Sun Z, Yan Z, Yao J, Beitler E, Zhu Y, Tour JM. Growth of graphene from solid carbon sources. Nature. 2010;**468**:549-552. DOI: 10.1038/nature09579

[42]  Chen Y, Gong XZ, Gai JG. Progress and challenges in transfer of large-area graphene films. Advancement of Science. 2016;**3**:1500343. DOI: 10.1002/advs.201500343

[43]  Sun J, Deng S, Guo W, Zhan Z, Deng J, Xu C, Fan X, Xu K, Guo W, Huang Y, Liu X. Electrochemical bubbling transfer of graphene using a polymer support with encapsulated air gap as permeation stopping layer. Journal of Nanomaterials. 2016;**2016**:1-7. DOI: 10.1155/2016/7024246

[44]  Cherian CT, Giustiniano F, Martin-Fernandez I, Andersen H, Balakrishnan J, Özyilmaz B. Bubble-free electrochemical delamination of CVD graphene films. Small. 2015;**11**:189-194. DOI: 10.1002/smll.201402024

[45]  Mafra D, Ming T, Kong J. Facial graphene transfer directly to target substrates with a reusable metal catalyst. Nanoscale. 2015;**7**:14807-14812. DOI: 10.1039/C5NR3892H

[46] Wang Y, Zheng Y, Xu X, Dubuisson E, Bao Q, Lu J, Loh KP. Electrochemical delamination of CVD-grown graphene film: Toward the recyclable use of copper catalyst. ACS Nano. 2011;**5**:9927-9933. DOI: 10.1021/nn203700w

[47] Terrones H, Lv R, Terrones M, Dresselhaus MS. The role of defects and doping in 2D graphene sheets and 1D nanoribbons. Reports on Progress in Physics. 2012;**75**:062501. DOI: 10.1088/0034-4885/75/6/062501

[48] Ambrosi A, Pumera M. The CVD graphene transfer procedure introduces metallic impurities which alter the graphene electrochemical properties. Nanoscale. 2014;**6**:472-476. DOI: 10.1039/C3NR05230C

[49] Lupina G, Kitsmann J, Costina I, Lukosius M, Wenger C, Wolff A, Vaziri S, Ostling M, Pasternak I, Krajewska A, Strupinski W, Kataria S, Gahoi A, Lemme MC, Ruhl G, Zoth G, Luxenhofer O, Mehr W. ACS Nano. 2015;**9**:4776-4785

[50] Ambrosi A, Bonanni A, Sofer Z, Pumera M. Large-scale quantification of CVD graphene surface coverage. Nanoscale. 2013;**5**:2379-2387. DOI: 10.1039/C3NR33824J

[51] Ismach A, Druzgalski C, Penwell S, Schwartzberg A, Zheng M, Javey A, Bokor J, Zhang Y. Direct chemical vapor deposition of graphene on dielectric surfaces. Nano Letters. 2010;**10**:1542-1548. DOI: 10.1021/nl9037714

[52] Levendorf MP, Ruiz-Vargas CS, Garg S, Park J. Transfer-free batch fabrication of single layer graphene transistors. Nano Letters. 2009;**9**:4479-4483. DOI: 10.1021/nl902790r

[53] Byun SJ, Lim H, Shin GY, Han TH, Oh SH, Ahn JH, Choi HC, Lee TW. Graphenes converted from polymers. Journal of Physical Chemistry Letters. 2011;**2**:493-497. DOI: 10.1021/jz200001g

[54] Lee CS, Baraton L, He Z, Maurice JL, Chaigneau M, Pribat D, Cojocaru CS. Dual graphene films growth process based on plasma-assisted chemical vapor deposition. Proceedings of SPIE. 2010;**7761**:77610P. DOI: 10.1117/12.861866

[55] Li X, Cai W, Colombo L, Ruoff RS. Evolution of graphene growth on Ni and cu by carbon isotope labeling. Nano Letters. 2009;**9**:4268-4272. DOI: 10.1021/nl902515k

[56] Li X, Zhu Y, Cai W, Borysiak M, Han B, Chen D, Piner RD, Colombo L, Ruoff RS. Transfer of large-area graphene films for high-performance transparent conductive electrodes. Nano Letters. 2009;**9**:4359-4363. DOI: 10.1021/nl902623y

[57] Rummeli MH, Bachmatiuk A, Scott A, Borrnert F, Warner JH, Hoffman V, Lin JH, Cuniberti G, Buchner B. Direct low-temperature nanographene CVD synthesis over a dielectric insulator. ACS Nano. 2010;**4**:4206-4210. DOI: 10.1021/nn100971s

[58] Zhang L, Shi Z, Wang Y, Yang R, Shi D, Zhang G. Ctalyst-free growth of nanographene films on various substrates. Nano Research. 2011;**4**:315-321. DOI: 10.1007/s12274-010-0086-5

[59] Chen J, Wen Y, Guo Y, Wu B, Huang L, Xue Y, Geng D, Wang D, Yu G, Liu Y. Oxygen-aided synthesis of polycrystalline graphene on silicon dioxide substrates. Journal of the American Chemical Society. 2011;**133**:17548-17551. DOI: 10.1021/ja2063633

[60]  Song HJ, Son M, Park C, Lim H, Levendorf MP, Tsen AW, Park J, Choi HC. Large scale metal-free synthesis of graphene on sapphire and transfer-free device fabrication. Nanoscale. 2012;**4**:3050-3054. DOI: 10.1039/C2NR30330B

[61]  Son M, Lim H, Hong M, Choi HC. Direct growth of graphene pad on exfoliated hexagonal boron nitride surface. Nanoscale. 2011;**3**:3089-3093. DOI: 10.1039/C1NR10504C

[62]  Teng PY, Lu CC, Akiyama-Hasegawa K, Lin YC, Yeh CH, Suenaga K, Chiu PW. Remote catalyzation for direct formation of graphene layers on oxides. Nano Letters. 2012;**12**:1379-1384. DOI: 10.1021/nl204024

[63]  Song X, Sun T, Yang J, Yu L, Wei D, Fang L, Lu B, Du C, Wei D. Direct growth of graphene films on 3D grating structural quartz substrates for high-performance pressure-sensitive sensors. ACS Applied Materials & Interfaces. 2016;**8**:16869-16875. DOI: 10.1021/acsami.6b04526

[64]  Ambrosi A, Pumera M. Electrochemistry at CVD grown multilayer graphene transferred onto flexible substrates. Journal of Physical Chemistry C. 2013;**117**:2053-2058. DOI: 10.1021/jp311739n

[65]  Loo AH, Ambrosi A, Bonanni A, Pumera M. CVD graphene based immuosensor. RSC Advances. 2014;**4**:23952-23956. DOI: 10.1039/c4ra03506b

[66]  Jiao T, Liu J, Wei D, Feng Y, Song X, Shi H, Jia S, Sun W, Du C. Composite transparent electrode of graphene nanowalls and silver nanowires on micropyramidal Si for high-efficiency schottky junction solar cells. ACS Applied Materials & Interfaces. 2015;**7**:20179-20183. DOI: 10.1021/acsami.5b05565

[67]  Steingrüber R, Ferstl M, Pilz W. Micro-optical elements fabricated by electron-beam lithography and dry etching technique using top conductive coatings. Microelectronic Eng. 2001;**57**:285-289. DOI: 10.1016/S0167-9317(01)00497-X

[68]  Mannsfeld SC, Tee BC, Stoltenberg RM, Chen CVH, Barman S, Muir BV, Sokolov AN, Reese C, Bao Z. Highly sensitive flexible pressure sensors with microstructured rubber dilectric layers. Nature Materials. 2010;**9**:859-864. DOI: 10.1038/nmat2834

[69]  Gogotsi Y. Nanomaterials Handbook: Florida, USA: CRC Press, Taylor & Francis Group; 2006. 800 p. https://www.crcpress.com/Nanomaterials-andbook/Gogotsi/p/book/9780849323089

[70]  Davis RF, Palmour III H, Porter RL. Emergent Process Methods for High-Technology Ceramics: New York, USA: Springer; 1984. 834 p. http://www.springer.com/gp/book/9781468482072

[71]  (a) Morosanu CE. Thin Films by Chemical Vapour Deposition: Amsterdam, Netherlands: Elsevier; 1990. p. 717. https://www.elsevier.com/books/thin-films-by-chemical-vapour-deposition/morosanu/978-0-444-98801-0. (b) Hess DW, Jensen KF, Anderson TJ. Chemical Vapor Deposition: A Chemical Engineering Perspective: Freund Publishing; 2011. DOI: 10.1515/REVCE.1985.3.2.97

[72]  Liu B, Tang DM, Sun C, Liu C, Ren W, Li F, Yu WJ, Yin LC, Zhang L, Jiang C. Importance of oxygen in the metal-free catalytic growth of single-walled carbon nanotubes from SiOx by a vapor-solid-solid mechanism. Journal of the American Chemical Society. 2010;**133**:197-199. DOI: 10.1021/ja107855q

[73] Chen Y, Zhang J. Diameter controlled growth of single-walled carbon nanotubes from SiO$_2$ nanoparticles. Carbon. 2011;**49**:3316-3324. DOI: 10.1016/j.carbon.2011.04.016

[74] Vishwakarma R, Rosmi MS, Takahashi K, Wakamatsu Y, Yaakob Y, Araby MI, Kalita G, Kitazawa M, Tanemura M. Transfer free graphene growth on SiO$_2$ substrate at 250 °C. Scientific Reports. 2017;**7**:43756. DOI: 10.1038/srep43756

[75] Saito K, Ogino T. Direct growth of graphene films on sapphire (0001) and (1120) surfaces by self-catalytic chemical vapor deposition. Journal of Physical Chemistry C. 2014;**118**:5523-5529. DOI: 10.1021/jp408126e

[76] Kang J, Shin D, Bae S, Hong BH. Graphene transfer: Key for applications. Nanoscale. 2012;**4**:5527-5537. DOI: 10.1039/C2NR3131K

[77] Kim YJ, Kim SJ, Jung MH, Choi KY, Bae S, Lee SK, Lee Y, Shin D, Lee B, Shin H. Low-temperature growth and direct transfer of graphene-graphitic carbon films on flexible plastic substrates. Nanotechnology. 2012;**23**:344016. DOI: 10.1088/0957-4484/23/34/344016

[78] Seo HK, Kim K, Min SY, Lee Y, Park CE, Raj R, Lee TW. Direct growth of graphene-dielectric bi-layer structure on device substrates from Si-based polymer. 2D Mater. 2017;**4**:024001. DOI: 10.1088/2053-1583/aa5408

[79] Yamada J, Ueda Y, Maruyama T, Naritsuka S. Direct growth of multilayer graphene by precipitation using W capping layer. Japanese Journal of Applied Physics. 2016;**55**:100302. DOI: 10.7567/JJAP.55.100302

[80] Su CY, Lu AY, Wu CY, Li YT, Liu KK, Zhang W, Lin SY, Juang ZY, Zhong YL, Chen FR. Direct formation of wafer scale graphene thin layers on insulating substrates by chemical vapor deposition. Nano Letters. 2011;**11**:3612-3616. DOI: 10.1021/nl201362n

[81] Peng Z, Yan Z, Sun Z, Tour JM. Direct growth of bilayer graphene on SiO$_2$ substrates by carbon diffusion through nickel. ACS Nano. 2011;**5**:8241-8247. DOI: 10.1021/nn202923y

[82] Min M, Seo S, Yoon Y, Cho K, Lee SM, Lee T, Lee H. Catalyst-free bottom-up growth of graphene nanofeatures along with molecules templates on dielectric substrates. Nanoscale. 2016;**8**:17022-17029. DOI: 10.1039/C6NR05657A

[83] Liu Z, Song L, Zhao S, Huang J, Ma L, Zhang J, Lou J, Ajayan PM. Direct growth of graphene/hexagonal boron nitride stacked layers. Nano Letters. 2011;**11**:2032-2037. DOI: 10.1021/nl200464j

[84] Trung PT, Delgado JC, Joucken F, Colomer JF, Hackens B, Raskin JP, Santos CN, Robert S. Direct growth of graphene on Si(111). Journal of Applied Physics. 2014;**115**:223704. DOI: 10.1063/1.4882181

[85] Xu SC, Man BY, Jiang SZ, Chen CS, Yang C, Liu M, Gao XG, Sun ZC, Zhang C. Direct synthesis of graphene on SiO$_2$ substrates by chemical vapor deposition. CrystEngComm. 2013;**15**:1840-1844. DOI: 10.1039/C3CE27029G

[86] Kwak J, Chu JH, Choi JK, Park SD, Go H, Kim SY, Park K, Kim SD, Kim YW, Yoon E. Near room-temperature synthesis of transfer-free graphene films. Nature Communications. 2012;**3**:645. DOI: 10.1038/ncomms1650

[87]  Bi H, Sun S, Huang F, Xie X, Jiang M. Direct growth of few-layer graphene films on $SiO_2$ substrates and their photovoltaic applications. Journal of Materials Chemistry. 2012;**22**:411-416. DOI: 10.1039/C1JM14778A

[88]  Kim H, Song I, Park C, Son M, Hong M, Kim Y, Kim JS, Shin HJ, Baik J, Choi HC. Copper-vapr-assisted chemical vapor deposition for high-quality and metal-free single-layer graphene on amorphous $SiO_2$ substrate. ACS Nano. 2013;**7**:6575-6582. DOI: 10.1021/nn402847w

[89]  Li J, Shen C, Que Y, Tian Y, Jiang L, Bao D, Wang Y, Du S, Gao HJ. Copper vapor-assisted growth of hexagonal graphene domains on silica islands. Applied Physics Letters. 2016;**109**:023106. DOI: 10.1063/1.4958872

[90]  Tang S, Wang H, Wang HS, Sun Q, Zhang X, Cong C, Xie H, Liu X, Zhou X, Huang F. Silane-catalysed fast growth of large single-crystalline graphene on hexagonal boron nitride. Nature Communications. 2015;**6**:6499. DOI: 10.1038/ncomms7499

[91]  Yang G, Kim HY, Jang S, Kim J. Versatile polymer-free graphene transfer method and applications. ACS Applied Materials & Interfaces. 2016;**8**:27115-27121. DOI: 10.1021/acsami.6b00681

[92]  Wang D, Tian H, Yang Y, Xie D, Ren TL, Zhang Y. Scalable and direct growth of graphene micro ribbons on dilectric substrates. Scientific Reports. 2013;**3**:1348. DOI: 10.1038/srep01348

[93]  Mehta R, Chugh S, Chen Z. Transfer-free multi-layer graphene as a diffusion barrier. Nanoscale. 2017;**9**:1827-1833. DOI: 10.1039/C6NR07637H

[94]  Muñoz R, Munuera C, Martínez J, Azpeitia J, Gómez-Aleixandre C, García-Hernández M. Low temperature metal free growth of graphene on insulating substrates by plasma assisted chemical vapor deposition. 2D Materials. 2016;**4**:015009. DOI: 10.1088/2053-1583/4/1/015009

[95]  Pang J, Mendes RG, Wrobel PS, Wlodarski MD, Ta HQ, Zhao L, Giebeler L, Trzebicka B, Gemming T, Fu L. Self-terminating confinement approach for large-area uniform monolayer directly over Si/SiOx by chemical vapor depostion. ACS Nano. 2017;**11**:1946. DOI: 10.1021/acsnano.6b08069

[96]  Sun J, Chen Y, Priydarshi MK, Chen Z, Bachmatiuk A, Zou Z, Chen Z, Song X, Gao Y, Rümmeli MH, Zhang Y, Liu Z. Nano Letters. 2015;**15**:5846-5854

[97]  Lee JH, Kim MS, Lim JY, Jung SH, Kang SG, shin HJ, Choi JY, Hwang SW, Whang D. Direct chemical vapor deposition-derived graphene glasses taegeting wide ranged applications. Applied Physics Letters 2016;**109**:053102. DOI: 10.1021/acs.nanolett.5b01936

[98]  Marchena M, Janner D, Chen TL, Finazzi V, Pruneri V. Low temperature direct growth of graphene patterns on flexible glass substrates catalysed by a sacrificial ultrathin Ni film. Opt. Mater. Express. 2016;**6**:2487-2057. DOI: 10.1364/OME.6.002487

[99]  Kalita G, Sugiura T, Wakamatsu Y, Hirano R, Tanemura M. Controlling the direct growth of graphene on an insulating substrate by the solid phase reaction of a polymer layer. RSC Advances. 2014;**4**:38450-38454. DOI: 10.1039/ C4RA05393A

[100] Sun J, Chen Y, Cai X, Ma B, Chen Z, Priydarshi MK, Chen K, Gao T, Song X, Ji Q, Guo X, Zou D, Zhang Y, Liu Z. Direct low-temperature synthesis of graphene on various glasses by plasma-enhanced chemical vapor deposition for versatile, cost-effective electrodes. Nano Research. 2015;**8**:3496-3504. DOI: 10.1007/s12274-015-0849-0

[101] Choi S, Lee H, Ghaffari R, Hyeon T, Kim DH. Recent advances in flexible and stretchable bio-electronic devices intergrated with nanomaterials. Advanced Materials. 2016;**28**:4203-4218. DOI: 10.1002/adma.201504150

[102] Hammock ML, Chortos A, Tee BCK, Tok JBH, Bao Z. 25[th] anniversary article: The evolution of electronic skin (e-skin): A brief history, design considerations, and recent progress. Adv. Mat. 2013;**25**:5997-6038. DOI: 10.1002/adma.201302240

[103] Kim H, Ahn JH. Graphene for flexible and wearable device applications. Carbon. 2017;**120**:244-257. DOI: 10.1016/j.carbon.2017.05.041

[104] Fan G, Zhu H, Wang K, Wei J, Li X, Shu Q, Guo N, Wu D. Graphene/silicon nanowire schottky junction for enhaced light harvesting. ACS Applied Materials & Interfaces. 2011;**3**:721-725. DOI: 10.1021/am1010354

[105] Wu T, Ding G, Shen H, Wang H, Sun L, Jiang D, Xie X, Jiang M. Triggering the continuous growth of graphene toward milimeter-sized grains. Advanced Functional Materials. 2013;**23**:198-203. DOI: 10.1002/adfm.201201577

[106] Yan Z, Lin J, Peng Z, Sun Z, Zhu Y, Li L, Xiang C, Samuel EL, Kittrell C, Tour JM. Toward the synthesis of wafer-scale single-crystal graphene on copper foils. ACS Nano. 2012;**6**:9110-9117. DOI: 10.1021/nn303352k

[107] Ma T, Ren W, Liu Z, Huang L, Ma LP, Ma X, Zhang Z, Peng LM, Cheng HM. Repeated growth-etching-regrowth for large-area defect-free single-crystal graphene by chemical vapor deposition. ACS Nano. 2014;**8**:12806-12813. DOI: 10.1021/nn506041t

[108] Zhang Z, Xu X, Qiu L, Wang S, Wu T, Ding F, Peng H, Liu K. The way towards ultrafast growth of single-crystal graphene on copper. Advancement of Science. 2017;**4**:1700087. DOI: 10.1002/advs.201700087

[109] Nguyen VL, Shin BG, Duong DL, Kim ST, Perello D, Lim YJ, Yuan QH, Ding F, Jeong HY, Shin HS, Lee SM, Chae SH, Vu QA, Lee SH, Lee YH. Seamless stitching of graphene domains on polished copper (111) foil. Advanced Materials. 2015;**27**:1376. DOI: 10.1002/adma.201404541

[110] Brown L, Lochocki EB, Avila J, Kim CJ, Ogawa Y, Havener RW, Kim DK, Monkman EJ, Shai DE, Wei HI, Levendorf MP, Asensio M, Shen KM, Park J. Polycrystalline graphene with single crystalline electronic structure. Nano Letters. 2014;**14**:5706. DOI: 10.1021/nl502445j

[111] Xue J, Sanchez-Yamagishi J, Bulmash D, Jacquod P, Deshpande A, Watanabe K, Taniguchi T, Jarillo-Herrero P, LeRoy BJ. Scanning tunnelling microscopy and spectroscopy of ultra-flat graphene on hexagonal boron nitride. Nature Materials. 2011;**10**:282-285. DOI: 10.1038/nmat2968